\IfFileExists{ieeeconf.cls}{%
  \documentclass[letterpaper,10pt,conference]{ieeeconf}%
}{%
  \documentclass[letterpaper,10pt,conference]{IEEEtran}%
}
\IEEEoverridecommandlockouts
\providecommand{\overrideIEEEmargins}{}
\usepackage{amsmath,amssymb,amsfonts,mathtools,bm}
\usepackage{graphicx}
\usepackage{cite}
\usepackage{booktabs}
\usepackage{microtype}
\usepackage{placeins}
\usepackage{balance}
\graphicspath{{figures/}}

\makeatletter
\@ifclassloaded{IEEEtran}{%
  \setlength{\textwidth}{7.0in}%
  \setlength{\columnsep}{0.2in}%
  \setlength{\oddsidemargin}{-0.25in}%
  \setlength{\evensidemargin}{-0.25in}%
  \setlength{\topmargin}{-0.25in}%
  \setlength{\textheight}{9.25in}%
}{}
\makeatother

\newcommand{\R}{\mathbb{R}}
\newcommand{\col}{\operatorname{col}}
\newcommand{\tr}{\operatorname{tr}}

\newcommand{\eps}{\varepsilon}

\title{\LARGE \bf
Extended Deterministic Learning under Local Finite-Time Excitation
}

\author{Chengzhi Yuan$^{1}$,  %
Mingxi Zhou$^{2}$, 
and Yuewei Fu$^{3}$
\thanks{This work is supported by the National Science Foundation (NSF) under Award 2238168 and by the Graduate School of Oceanography, University of Rhode Island.} 
\thanks{$^{1}$C. Yuan is with the Department of Mechanical, Industrial and Systems Engineering,
University of Rhode Island, Kingston, RI 02881, USA. Email: cyuan@uri.edu}%
\thanks{$^{2}$M. Zhou is with the Graduate School of Oceanography, University of Rhode Island, Narragansett, RI 02882, USA. Email: mzhou@uri.edu}%
\thanks{$^{3}$Y. Fu is with the Department of Ocean Engineering, University of Rhode Island, Narragansett, RI 02882, USA. Email: yweifu@uri.edu}%
}

\begin{document}
\maketitle
\thispagestyle{empty}
\pagestyle{empty}

\begin{abstract}
This paper develops an extension of classical deterministic learning (CDL) for the regulation of uncertain nonlinear systems, referred to as \emph{Extended Deterministic Learning} (EDL). CDL relies on periodic or recurrent closed-loop trajectories to ensure that the locally active radial-basis-function (RBF) neurons satisfy a partial persistence-of-excitation (PE) condition. During regulation, however, the state approaches a constant operating point and continued excitation disappears. EDL preserves the local-learning mechanism of CDL while enabling locally accurate learning under a \emph{local finite-time excitation} (local FTE) condition, which requires neither persistent excitation nor recurrent system trajectories. Specifically, an adaptive RBF controller is combined with filtered finite-time data storage. Under local FTE, the resulting scheme guarantees that all closed-loop signals are uniformly ultimately bounded, the regulation state converges exponentially to a small neighborhood of the desired operating point, and the neural weights associated with the locally active RBFs converge exponentially to small neighborhoods of their ideal values. Consequently, the unknown nonlinear dynamics are learned locally and accurately over the experienced region without requiring persistent excitation after the finite learning interval. A numerical example illustrates the improvement in local learning achieved by EDL over CDL.
\end{abstract}

\section{Introduction}
Adaptive neural control provides a systematic framework for compensating uncertain nonlinear dynamics while maintaining closed-loop stability. In particular, linearly parameterized neural approximators such as Gaussian radial-basis-function (RBF) networks are attractive because their weights can be updated by Lyapunov-based laws while the basis functions retain a spatially localized structure \cite{Narendra1990, Sanner1992}.

Deterministic learning (DL) goes beyond control by asking whether the unknown nonlinear dynamics experienced during feedback control can be identified and retained. In the classical theory of DL, a periodic or recurrent trajectory repeatedly activates a local group of RBF neurons \cite{WangHill2006}. Because of RBF localization, the corresponding local regressor subvector satisfies a partial persistence-of-excitation (PE) condition, even though the complete RBF network does not need to be persistently excited \cite{WangHill2007}. The resulting local parameter convergence yields a locally accurate, time-invariant neural representation of the system dynamics along the recurrent trajectory \cite{WangChen2009}. This local-learning interpretation is a defining feature of DL and is developed systematically in \cite{WangHillBook}. Recent studies have extended the DL framework to sampled-data systems \cite{ZhangWuHuWang2022}, knowledge fusion across learned regions \cite{WuHuZhuZhangXuWang2024}, and broader nonlinear control settings \cite{YangZhangWang2026}.

The recurrent-trajectory requirement, however, can become restrictive for regulation tasks. When the objective is to drive the state to a constant operating point, the useful transient trajectory is finite and nonrecurrent; the locally active RBFs may be sufficiently excited only for a limited time before the state settles. Related adaptive-control research has shown that parameter convergence can sometimes be recovered from stored finite data rather than classical PE. Concurrent learning is an early example of this idea \cite{Chowdhary2010}. Composite learning subsequently used historical or filtered data to relax PE requirements \cite{PanYu2016}, while finite-excitation conditions were established explicitly for adaptive convergence \cite{Cho2018}. Integral concurrent learning further showed that informative finite intervals can support parameter convergence without indefinite excitation \cite{Parikh2019}. Related cooperative finite-time-excitation results extended this principle to networked adaptive systems \cite{Yuan2021}. These results motivate the use of finite information, but they do not by themselves resolve the central DL question considered here: how to preserve the \emph{localized} learning property of classical DL when the experienced trajectory is nonrecurrent.

This paper addresses that question by developing \emph{Extended Deterministic Learning} (EDL) for nonlinear regulation. The key concept is \emph{local finite-time excitation} (local FTE). Instead of requiring recurrent partial PE for all future time, EDL requires only a finite interval over which the corresponding local RBF regressor subvector is sufficiently excited. No positive-definiteness requirement is imposed on the complete information matrix. Thus, remote RBF neurons need not be excited, and the full network may remain rank deficient. This point is essential: EDL replaces the \emph{temporal requirement} of recurrent partial PE by one finite interval of local excitation without abandoning the \emph{spatial locality} on which classical DL is built.

The proposed EDL framework consists of three tightly connected components: an adaptive neural controller for nonlinear regulation, a set of stable filters that transform the available closed-loop signals into a form suitable for data-based learning without requiring state derivatives, and a memory-enhanced neural learner that combines current adaptive information with finite transient data retained from the experienced trajectory. The controller provides the closed-loop regulation behavior that generates the finite learning experience, the filters extract and organize the information contained in this experience, and the learner preserves and reuses the resulting information after the transient excitation has disappeared. A rigorous Lyapunov-based analysis is then developed to establish uniform ultimate boundedness of the complete adaptive closed-loop system, exponential convergence of the regulation state to a small neighborhood of the desired operating point, exponential convergence of the locally active neural weights under local FTE, and locally accurate learning of the nonlinear dynamics encountered along the experienced region. Finally, a numerical example with a nonrecurrent regulation trajectory is used to illustrate the proposed mechanism and to demonstrate that EDL can retain an accurate local model of the experienced nonlinear dynamics, whereas the adaptive neural controller under the classical DL mechanism can achieve regulation but does not preserve comparable learned knowledge.

The remainder of this paper is organized as follows. Section~II reviews notation and the classical deterministic-learning (CDL) mechanism and then states the regulation-learning problem. Section~III develops the EDL framework, including the filtered regression, finite-data memory, local FTE condition, adaptive regulator, convergence analysis, and implementation remarks. Section~IV presents the numerical study and comparison with the CDL method. Section~V concludes the paper.

\section{Preliminaries and Problem Statement}
\subsection{Notation}\label{sec.notation}
For a vector $x$, $\|x\|$ denotes the Euclidean norm. For a matrix $X$, $\|X\|$ and $\|X\|_F$ denote the induced 2-norm and Frobenius norm, respectively, and $\tr(X)$ denotes the trace. For a symmetric matrix $X$, $X\succeq0$ ($X\succ0$) denotes positive semidefiniteness (positive definiteness), and $\lambda_{\min}(X)$ and $\lambda_{\max}(X)$ denote its minimum and maximum eigenvalues. The notation $\col(x_1,\ldots,x_r)$ denotes column stacking, $I_r$ is the $r\times r$ identity matrix, and $\R^r$ is the $r$-dimensional Euclidean space. A hat denotes an estimate and a tilde denotes the corresponding estimation error, e.g., $\widetilde W=\widehat W-W^*$. The index set $\zeta$ will denote the locally active RBF neurons and $\bar\zeta$ the remaining neurons, both used only in the analysis.

\subsection{Classical Deterministic Learning}
The theory of classical deterministic learning (CDL) was developed to identify nonlinear dynamical systems undergoing periodic or recurrent motion \cite{WangHill2007}. Consider
\begin{equation*} \dot x=F(x;p),
 \qquad x\in\R^n,
\end{equation*}
where $p$ is a constant parameter vector and $F(x;p)=\col(f_1(x;p),\ldots,f_n(x;p))$ is unknown and continuous. Suppose that, from a given initial condition, the system generates a bounded periodic or periodic-like recurrent trajectory $\varphi(x_0;p)$. The objective of DL is to identify the unknown dynamics $F(x;p)$ locally along this recurrent trajectory.

For each component $f_i(x;p)$, the dynamical localized RBF identifier used in classical DL has the form
\begin{equation*} \dot{\widehat x}_i=-a_i(\widehat x_i-x_i)+\widehat W_i^T S_i(x),
 \qquad a_i>0,
\end{equation*}
where $S_i(x)$ is a localized RBF vector. The RBF network is linear in its weights and, with sufficiently many nodes placed on a regular lattice over a compact region, has the universal approximation property:
\begin{equation*} f_i(x;p)=W_i^{*T}S_i(x)+\eps_i(x).
\end{equation*}
The localization property of RBF networks is essential: along a bounded trajectory, only a reduced subvector $S_{\zeta_i}(x)$ associated with centers near the trajectory is appreciably activated, so that 
\begin{equation*} f_i(x;p)=W_{\zeta_i}^{*T}S_{\zeta_i}(x)+\eps_{\zeta_i}(x),
\end{equation*}
with a small bounded localized approximation error $\eps_{\zeta_i}(x)$.

The central result of CDL is that recurrence supplies the required excitation automatically. For a periodic or recurrent trajectory that remains inside the RBF lattice, the locally active regressor subvector is persistently exciting; equivalently, there exist $T_0>0$ and $\alpha_i>0$ such that
\begin{equation*} \int_t^{t+T_0}S_{\zeta_i}(x(\tau))S_{\zeta_i}^T(x(\tau))d\tau
 \succeq \alpha_i I
\end{equation*}
for all sufficiently large $t$. This is the classical \emph{partial PE} property, i.e., PE is required only for the RBFs locally activated by the recurrent orbit, not for the complete network.

The neural weights are updated by the Lyapunov-based law
\begin{equation*} \dot{\widehat W}_i
 =-\Gamma_i S_i(x)\widetilde x_i
 -\sigma_i\Gamma_i\widehat W_i,
 \qquad \widetilde x_i=\widehat x_i-x_i,
\end{equation*}
where $\Gamma_i\succ0$ and $\sigma_i>0$ is small. Owing to partial PE of $S_{\zeta_i}$, the nominal local identification-error dynamics are exponentially stable. Consequently, the state-identification error and the locally active parameter errors converge exponentially to small neighborhoods of zero, and the RBF network gives a locally accurate approximation of the unknown dynamics along the recurrent trajectory. After the transient, a constant weight vector can be obtained by averaging the converged estimates over a post-transient time segment; the resulting constant RBF network provides a time-invariant, locally accurate representation of the learned dynamics \cite{WangHill2007}.
This recurrent-partial-PE mechanism is the CDL result that the present work replaces with local finite-time excitation.

\subsection{Problem Statement}
This paper considers the following matched-uncertainty nonlinear system:
\begin{equation}
 \dot x=Ax+B\left[u+f(x)\right],
 \qquad x\in\R^n,\quad u\in\R^m,
 \label{eq:plant}
\end{equation}
where $x$ and $u$ are the system state and input, respectively, $A$ and $B$ are known constant matrices of compatible dimensions, and $f:\R^n\to\R^m$ is a smooth matched nonlinear uncertainty. 

The assumptions are standard: \emph{(A1)} $(A,B)$ is stabilizable and $B\in\R^{n\times m}$ has full column rank; and \emph{(A2)} over a compact design region $\Omega$, $f(x)$ is smooth and admits a complete localized Gaussian-RBF approximation: $f(x)=W^{*T}S(x)+\eps(x)$ with constant $W^*$ and $\|\eps(x)\|\le\bar\eps$.

The objective is to design an adaptive learning controller with RBF networks to simultaneously regulate $x$ and learn the unknown nonlinear dynamics encountered during a finite transient. Specifically, under finite-time excitation of the locally active RBFs, the corresponding local weights should converge to small neighborhoods of their associated ideal values so that $f(x)$ can be learned locally and accurately along the experienced trajectory and over the associated region. 
In contrast to classical DL, the regulation trajectory is not assumed to be periodic or recurrent and may converge to a constant operating point. Thus, the learning objective is achieved without requiring partial PE or persistent recurrence of system trajectories. The precise local finite-time excitation (FTE) condition will be introduced in the next section. 

\section{Main Results}
For the plant \eqref{eq:plant}, the Gaussian RBF approximation described in Section~II-B gives
\begin{equation}
 f(x)=W^{*T}S(x)+\eps(x),\qquad \|\eps(x)\|\le\bar\eps.
 \label{eq:rbf}
\end{equation}
where $W^*\in\R^{q\times m}$ is the ideal weight matrix, $S(x)$ is the RBF regressor vector of compatible dimensions, and $\eps(x)$ is the approximation error, bounded by a small constant $\bar\eps>0$. 

Because $f(x)$ is not assumed to be measurable, we first construct filtered signals for the controller and learner. Choose $\lambda_f>0$ and introduce
\begin{equation}
\begin{aligned}
 \dot x_f&=-\lambda_f x_f+x,\\
 \dot u_f&=-\lambda_f u_f+u,\\
 \dot S_f&=-\lambda_f S_f+S(x),
\end{aligned}
\label{eq:filters}
\end{equation}
with zero initial conditions. With $B^\dagger :=(B^TB)^{-1}B^T$, define
\begin{equation}
\begin{aligned}
 y=B^\dagger\big[&x-e^{-\lambda_ft}x(0)-\lambda_fx_f\\
 &-Ax_f-Bu_f\big].
\end{aligned}
\label{eq:y}
\end{equation}
To make the regression residual explicit, apply the same stable filter $1/(s+\lambda_f)$ to \eqref{eq:plant}. Since
\[
 x_f(t)=\int_0^t e^{-\lambda_f(t-\tau)}x(\tau)d\tau,
\]
integration by parts gives
\[
\begin{aligned}
 \frac{1}{s+\lambda_f}\dot x
 &=\int_0^t e^{-\lambda_f(t-\tau)}\dot x(\tau)d\tau\\
 &=x-e^{-\lambda_ft}x(0)-\lambda_fx_f.
\end{aligned}
\]
Therefore, filtering \eqref{eq:plant}, together with \eqref{eq:filters}, yields
\[
 x-e^{-\lambda_ft}x(0)-\lambda_fx_f
 =Ax_f+Bu_f+B\big(W^{*T}S_f+d\big),
\]
where
\[
 d(t)=\int_{0}^{t}e^{-\lambda_f(t-\tau)}\eps(x(\tau))\,d\tau.
\]
Moving the known terms to the left and premultiplying by $B^\dagger$ gives $y=B^\dagger B(W^{*T}S_f+d)$. Since $B^\dagger B=I_m$, the definition of $y$ in \eqref{eq:y} yields
\begin{equation}
 y=W^{*T}S_f+d.
 \label{eq:regression}
\end{equation}
Moreover, because $\|\eps(x)\|\le\bar\eps$, we have
\[
 \|d(t)\|\le \int_0^t e^{-\lambda_f(t-\tau)}\bar\eps\,d\tau
 =\frac{\bar\eps}{\lambda_f}\big(1-e^{-\lambda_ft}\big)
 \le\frac{\bar\eps}{\lambda_f}\triangleq\bar d.
\]

Over a finite interval $[t_a,t_b]$ with $t_b>t_a\geq 0$, initialize $M(t_a)=0$ and $Q(t_a)=0$ and accumulate the full-dimensional information matrices online according to
\begin{equation}
 \dot M=S_fS_f^T,\qquad \dot Q=S_fy^T,
 \qquad t\in[t_a,t_b].
 \label{eq:MQdot}
\end{equation}
Equivalently, for $t\in[t_a,t_b]$,
\begin{equation}
 M(t)=\int_{t_a}^{t}S_f(\tau)S_f^T(\tau)d\tau,\qquad
 Q(t)=\int_{t_a}^{t}S_f(\tau)y^T(\tau)d\tau.
 \label{eq:MQ}
\end{equation}
The memory term uses the running matrices $M(t)$ and $Q(t)$ from the beginning of the interval. At $t=t_b$, data collection stops and the matrices are frozen,
\begin{equation*} M(t)=M(t_b),\qquad Q(t)=Q(t_b),\qquad t\ge t_b.
\end{equation*}
From \eqref{eq:regression}, define
\begin{equation*} D(t)=\int_{t_a}^{\min\{t,t_b\}}S_f(\tau)d^T(\tau)d\tau,
\end{equation*}
so that
\begin{equation}
 Q(t)=M(t)W^*+D(t),
 \label{eq:Qrelation}
\end{equation}
Let
\begin{equation*} \bar S_f\triangleq\sup_{\tau\in[t_a,t_b]}\|S_f(\tau)\|,
 \qquad T_e\triangleq t_b-t_a.
\end{equation*}
Since $\|d(t)\|\le\bar d=\bar\eps/\lambda_f$, the accumulated regression residual satisfies
\begin{align*} \|D(t)\|_F
 &\le \int_{t_a}^{\min\{t,t_b\}}
       \|S_f(\tau)\|\,\|d(\tau)\|d\tau \\
 &\le T_e\bar S_f\bar d
 =\frac{T_e\bar S_f\bar\eps}{\lambda_f}
 \triangleq \bar D.
\end{align*}
The quantity $\bar D$ is an explicit upper bound on the finite accumulated effect of the filtered RBF approximation error in the stored regression.

Let the finite closed-loop trajectory remain in a compact local region $\Omega_L\subset\Omega$. For the subsequent analysis, we use the localized decomposition $S=\col(S_{\zeta},S_{\bar\zeta})$ and the associated filtered partition $S_f=\col(S_{f,\zeta},S_{f,\bar\zeta})$, so $M(t)$ in \eqref{eq:MQ} can be partitioned as $M = \begin{bmatrix}M_{\zeta\zeta} & M_{\zeta\bar\zeta}\\
M_{\bar\zeta\zeta} & M_{\bar\zeta\bar\zeta}\end{bmatrix}$ with 
\begin{equation*} M_{\zeta\zeta}=\int_{t_a}^{t_b}S_{f,\zeta}S_{f,\zeta}^T d\tau,
 \qquad
 M_{\zeta\bar\zeta}=\int_{t_a}^{t_b}S_{f,\zeta}S_{f,\bar\zeta}^T d\tau.
\end{equation*}
As defined in Section~\ref{sec.notation}, $\zeta$ denotes the locally active RBF neurons along the system's state trajectory and $\bar\zeta$ the remaining neurons. The stable filter preserves RBF localization in the standard practical sense: over $[t_a,t_b]$ there exist finite $\bar S_{f,\zeta}$ and small $\delta_{S_f}>0$ such that
\begin{equation*} \|S_{f,\zeta}(t)\|\le\bar S_{f,\zeta},
 \qquad
 \|S_{f,\bar\zeta}(t)\|\le\delta_{S_f},
\end{equation*}
which implies
\begin{equation}
 \|M_{\zeta\bar\zeta}\|
 \le T_e\bar S_{f,\zeta}\delta_{S_f}.
 \label{eq:cross-small}
\end{equation}

We now introduce the local FTE condition. 

\textbf{Definition 1 (local FTE):}
The finite trajectory on $[t_a,t_b]$ is said to satisfy \emph{local finite-time excitation} (local FTE) if there exists $\alpha_\zeta>0$ such that
\begin{equation}
 {M_{\zeta\zeta}\succeq\alpha_\zeta I_{q_\zeta}.}
 \label{eq:local-fte}
\end{equation}

Note that no positive-definiteness condition is imposed on the complete matrix $M$. Hence, remote RBF neurons need not be excited, consistent with the local-learning property of classical DL.

To design the controller and learner, let $P=P^T\succ0$ satisfy
\begin{equation}
 A^TP+PA-PBB^TP+I=0,
 \label{eq:ARE}
\end{equation}
and select
\begin{equation}
 K=-k_cB^TP,\qquad k_c>\frac12.
 \label{eq:K}
\end{equation}
The adaptive neural controller is
\begin{equation}
 {u=Kx-\widehat W^TS(x).}
 \label{eq:controller}
\end{equation}
The adaptive learning law is
\begin{equation}
{\dot{\widehat W}=\Gamma\left[
 S(x)x^TPB-\sigma\widehat W
 -\kappa\big(M(t)\widehat W-Q(t)\big)\right],}
 \label{eq:learning-law}
\end{equation}
where $\Gamma=\gamma I_q\succ0$, $\sigma>0$ is a small leakage gain, and $\kappa>0$. Thus the complete RBF network and the memory term (i.e., $M\widehat W-Q$) are used from the beginning of data collection. During $[t_a,t_b]$, the memory matrices $(M,Q)$ evolve online according to \eqref{eq:MQdot}; after $t_b$, they remain frozen while the same memory term continues to drive learning.

\textbf{Theorem 1 (EDL under local FTE):}
Consider the plant \eqref{eq:plant} under controller \eqref{eq:controller} and learner \eqref{eq:learning-law} with the filters in \eqref{eq:filters}, the regression output in \eqref{eq:y}, and the memory updates in \eqref{eq:MQdot}. Suppose that the closed-loop trajectory generated during $[t_a,t_b]$ remains in a compact region $\Omega_L\subset\Omega$, and that the RBF subvector satisfies the local FTE condition \eqref{eq:local-fte}. Then: (i) all signals of the closed-loop system are uniformly ultimately bounded; (ii) the system state $x$ converges exponentially to a small neighborhood of the origin; (iii) the associated local neural weight estimates converge exponentially to small neighborhoods of their ideal values; and (iv) $\widehat W^TS(x)$ provides a locally accurate approximation of the unknown nonlinear dynamics on $\Omega_L$. 

\emph{Proof:}
The proof follows the standard CDL sequence \cite{WangHill2006,WangHillBook}; EDL changes only the excitation mechanism in the final local-learning step, where recurrent partial PE is replaced by local FTE of the stored filtered regression.

\emph{Step 1: boundedness of the complete adaptive closed-loop system.}
Let
\begin{equation*} \widetilde W=\widehat W-W^*. 
\end{equation*}
Using \eqref{eq:plant}, \eqref{eq:rbf}, and \eqref{eq:controller},
\begin{equation*} \dot x=(A+BK)x-B\widetilde W^TS(x)+B\eps(x).
\end{equation*}
From \eqref{eq:ARE} and \eqref{eq:K},
\begin{equation*} (A+BK)^TP+P(A+BK)
 =-I-(2k_c-1)PBB^TP.
\end{equation*}
Consider the following Lyapunov function for the closed-loop system:
\begin{equation*} V=x^TPx+\tr(\widetilde W^T\Gamma^{-1}\widetilde W).
\end{equation*}
From \eqref{eq:Qrelation} and \eqref{eq:learning-law},
\begin{align*} \dot V
 &=-\|x\|^2-(2k_c-1)\|B^TPx\|^2+2x^TPB\eps \\
 &\quad-2\sigma\tr(\widetilde W^T\widehat W)
 -2\kappa\tr(\widetilde W^TM(t)\widetilde W)\\
 &\quad+2\kappa\tr(\widetilde W^TD(t)).
\end{align*}
Furthermore,
\begin{equation*} -2\tr(\widetilde W^T\widehat W)
 \le-\|\widetilde W\|_F^2+\|W^*\|_F^2.
\end{equation*}
Young's inequality gives
\begin{align*} 2x^TPB\eps&\le\frac{2k_c-1}{2}\|B^TPx\|^2
 +\frac{2\bar\eps^2}{2k_c-1},\\
 2\kappa\tr(\widetilde W^TD(t))&\le\frac{\sigma}{2}\|\widetilde W\|_F^2
 +\frac{2\kappa^2}{\sigma}\bar D^2.
\end{align*}
Hence
\begin{equation*} \dot V\le-\|x\|^2-\frac{2k_c-1}{2}\|B^TPx\|^2
 -\frac{\sigma}{2}\|\widetilde W\|_F^2+C_0,
\end{equation*}
where
\begin{equation*} C_0=\frac{2\bar\eps^2}{2k_c-1}+\sigma\|W^*\|_F^2
 +\frac{2\kappa^2}{\sigma}\bar D^2.
\end{equation*}
Since $k_c>1/2$, the term $-\frac{2k_c-1}{2}\|B^TPx\|^2$ is nonpositive and may be dropped when forming an upper bound on $\dot V$. Moreover, with $\bar c_V\triangleq\max\{\lambda_{\max}(P),\lambda_{\max}(\Gamma^{-1})\}$, we have
\[
 V\le \bar c_V\big(\|x\|^2+\|\widetilde W\|_F^2\big).
\]
Therefore,
\begin{equation*} \dot V\le-\alpha_VV+C_0,
 \qquad
 \alpha_V=\frac{\min\{1,\sigma/2\}}{\bar c_V}>0.
\end{equation*}
Thus
\begin{equation*} V(t)\le e^{-\alpha_Vt}V(0)
 +\frac{C_0}{\alpha_V}(1-e^{-\alpha_Vt}),
\end{equation*}
which establishes uniform ultimate boundedness of all the closed-loop signals.

\emph{Step 2: convergence of the regulation state.}
From Step 1, let $\|\widetilde W(t)\|_F\le\bar W$ and $\|S(x)\|\le s^*$ on $\Omega$. For $V_x=x^TPx$,
\begin{equation*} \dot V_x\le-\|x\|^2-(2k_c-1)\|B^TPx\|^2
 +2\|B^TPx\|\bar\delta,
\end{equation*}
where $\bar\delta=\bar Ws^*+\bar\eps$. Therefore,
\begin{equation*} \dot V_x\le-\|x\|^2-\frac{2k_c-1}{2}\|B^TPx\|^2+C_x,
 \qquad
 C_x=\frac{2\bar\delta^2}{2k_c-1}.
\end{equation*}
Since $V_x\le\lambda_{\max}(P)\|x\|^2$,
\begin{equation*} \dot V_x\le-\alpha_xV_x+C_x,
 \qquad \alpha_x=\lambda_{\max}^{-1}(P),
\end{equation*}
so $x$ converges exponentially to a compact neighborhood of the origin.

\emph{Step 3: EDL of the locally active RBF weights under local FTE.} On $\Omega_L$, recall that 
\begin{equation*} S=\col(S_\zeta,S_{\bar\zeta}),\qquad
 \widehat W=\col(\widehat W_\zeta,\widehat W_{\bar\zeta}).
\end{equation*}
Using the local RBF representation,
\begin{equation*} \dot x=(A+BK)x-B\widetilde W_\zeta^TS_\zeta(x)
 +B\eps_\zeta'(x),
\end{equation*}
where
\begin{equation*} \eps_\zeta'=\eps_\zeta-\widehat W_{\bar\zeta}^TS_{\bar\zeta}.
\end{equation*}
Step 1 and Gaussian localization imply $\|\eps_\zeta'\|\le\bar\eps_\zeta'$ on $\Omega_L$. After $t_b$, the memory matrices are frozen. Taking the $\zeta$ rows of the complete adaptive law yields
\begin{equation*}\begin{aligned}
 \dot{\widetilde W}_\zeta=\Gamma_\zeta\big[&S_\zeta x^TPB -\sigma\widetilde W_\zeta
 -\kappa M_{\zeta\zeta}\widetilde W_\zeta+\Delta_\zeta\big],
\end{aligned}
\end{equation*}
where
\begin{equation*} \Delta_\zeta=-\sigma W_\zeta^*
 -\kappa(M_{\zeta\bar\zeta}\widetilde W_{\bar\zeta}-D_\zeta).
\end{equation*}
By Step 1 and \eqref{eq:cross-small}, $\|\Delta_\zeta\|_F\le\bar\Delta_\zeta$ for some small finite bound $\bar\Delta_\zeta$.

Consider the Lyapunov function for the local adaptive subsystem:
\begin{equation*} V_L=x^TPx+\tr(\widetilde W_\zeta^T
 \Gamma_\zeta^{-1}\widetilde W_\zeta).
\end{equation*}
By Definition 1, we have
\begin{align*} \dot V_L
 &\le-\|x\|^2-(2k_c-1)\|B^TPx\|^2
 -2\lambda_\zeta\|\widetilde W_\zeta\|_F^2\\
 &\quad+2\|B^TPx\|\bar\eps_\zeta'
 +2\|\widetilde W_\zeta\|_F\bar\Delta_\zeta,
\end{align*}
where $\lambda_\zeta=\sigma+\kappa\alpha_\zeta>0$. Therefore,
\begin{equation*} \dot V_L\le-\|x\|^2-\frac{2k_c-1}{2}\|B^TPx\|^2
 -\lambda_\zeta\|\widetilde W_\zeta\|_F^2+C_L,
\end{equation*}
with
\begin{equation*} C_L=\frac{2\bar\eps_\zeta'^2}{2k_c-1}
 +\frac{\bar\Delta_\zeta^2}{\lambda_\zeta}.
\end{equation*}
Let $\bar c_L=\max\{\lambda_{\max}(P),\lambda_{\max}(\Gamma_\zeta^{-1})\}$ and
\begin{equation*} \alpha_L=\frac{\min\{1,\lambda_\zeta\}}{\bar c_L}>0.
\end{equation*}
Then
\begin{equation*} \dot V_L\le-\alpha_LV_L+C_L,
\end{equation*}
which gives
\begin{equation}
 V_L(t)\le e^{-\alpha_L(t-t_b)}V_L(t_b)
 +\frac{C_L}{\alpha_L}\left(1-e^{-\alpha_L(t-t_b)}\right).
 \label{eq:VLexplicit}
\end{equation}
Thus the locally active neural weights converge exponentially to small neighborhoods of their ideal values.

Finally, for $x\in\Omega_L$,
\begin{equation*}\begin{aligned}
 f(x)-\widehat W^TS(x)
 =&-\widetilde W_\zeta^TS_\zeta(x)
 -\widehat W_{\bar\zeta}^TS_{\bar\zeta}(x)\\
 &+\eps_\zeta(x).
\end{aligned}
\end{equation*}
The first term is small by \eqref{eq:VLexplicit}; the second is small because the remote Gaussian RBFs are small on $\Omega_L$ and the complete weights are bounded from Step 1; and the last term is the local approximation residual. Hence, the complete RBF network provides a locally accurate approximation of the unknown dynamics over the experienced region. This establishes all statements of the theorem. \hfill$\square$

\textbf{Remark 1:} The local objects $S_\zeta$, $S_{f,\zeta}$, and $M_{\zeta\zeta}$ are introduced only to state and prove the learning property. The actual controller, filters, stored matrices $M,Q$, and adaptive law are full dimensional. The only role of local FTE is to replace the partial-PE condition in the final local-learning step of classical DL; this replacement defines the proposed EDL framework.

\textbf{Remark 2:} The theorem requires only the existence of a finite interval $[t_a,t_b]$ that provides local FTE; the terminal time $t_b$ need not be known a priori. In implementation, the memory term may be active from the beginning of the transient while the freezing time is selected online. One convenient leverage/novelty index is
\begin{equation}
 \nu(t)=S_f^T(t)\left[M(t)+\delta I\right]^{-1}S_f(t),\qquad \delta>0,
 \label{eq:nu-remark}
\end{equation}
which measures how much new regressor information the current sample contributes relative to the information already stored in $M(t)$. The matrices $M,Q$ may be frozen once $\nu(t)$ stays below a prescribed threshold for a dwell time after an initial transient. The theory above is unchanged; the resulting data-dependent stopping time only needs to satisfy the local-FTE condition $M_{\zeta\zeta}(t_b)\succ0$.

\textbf{Remark 3:} For online implementation, one integrates the three filters in \eqref{eq:filters}, uses $S_f$ directly, forms the measurable filtered regression output $y$ from \eqref{eq:y}, updates the memory during data collection according to \eqref{eq:MQdot}, applies the controller \eqref{eq:controller}, and updates the neural weight estimate using \eqref{eq:learning-law}. Equivalently, the memory-writing dynamics can be summarized as
\begin{equation*} (\dot M,\dot Q)=
 \begin{cases}
 (S_fS_f^T,\,S_fy^T), & \text{while collecting data},\\
 (0,0), & \text{after the memory is frozen}.
 \end{cases}
\end{equation*}
The transition from data collection to frozen memory can follow the online stopping rule in Remark~2, using $\nu(t)$ in \eqref{eq:nu-remark}. 
The controller and weight update continue to run after freezing, with the fixed matrices $M(t_b),Q(t_b)$.

\section{Illustrative Example}
Consider the following second-order nonlinear system of the form \eqref{eq:plant}:
\begin{equation*} \dot x_1=x_2,\qquad
 \dot x_2=u+f(x_1,x_2),
\end{equation*}
with $x(0)=[2.15\;-1.10]^T$. The unknown plant nonlinearity is specified as
\begin{equation*} f(x_1,x_2)=\frac{0.3}{1+x_1^2}+0.12\sin x_1+0.04x_2^2.
\end{equation*}
The implemented RBF network contains $63$ Gaussian RBF neurons with common width $\eta=0.28$. Their centers form the uniform $9\times7$ grid
\begin{equation*} c_{ij}=\begin{bmatrix}-1+0.5(i-1)\\-1.5+0.5(j-1)\end{bmatrix},
 \quad i=1,\ldots,9,\ j=1,\ldots,7.
\end{equation*}

For the controller, solving \eqref{eq:ARE} gives
$
 P=\begin{bmatrix}1.7321&1\\1&1.7321\end{bmatrix} $ and $
 K=-2.2B^TP=\begin{bmatrix}-2.2&-3.8105\end{bmatrix}$, 
so the implemented adaptive controller is
\begin{equation*} u=-2.2x_1-3.8105x_2-\widehat W^TS(x).
\end{equation*}
The learner uses $\lambda_f=1.2$, $\Gamma=30I_{63}$, $\sigma=2\times10^{-3}$, and $\kappa=20$, and therefore
\begin{equation*} \dot{\widehat W}=30\!\left[S(x)x^TPB-2\times10^{-3}\widehat W-20(M\widehat W-Q)\right].
\end{equation*}
The initial conditions are $\widehat W(0)=0$, $x_f(0)=u_f(0)=S_f(0)=0$, and $M(0)=Q(0)=0$. The memory term is active from $t=0$. Instead of prescribing $t_b$, the online criterion in Remark~2 is implemented with $\delta=10^{-4}$, $\epsilon_\nu=0.8$, $T_{\min}=5$ s, and dwell time $T_d=1$ s. It automatically freezes the memory at
\begin{equation*} t_b=9.927~\text{s},
\end{equation*}
after which the frozen information remains in the learning law. The simulation horizon is $100$ s.

\begin{figure*}[!t]
\centering
\begin{minipage}{0.31\textwidth}
\centering
\includegraphics[width=\linewidth]{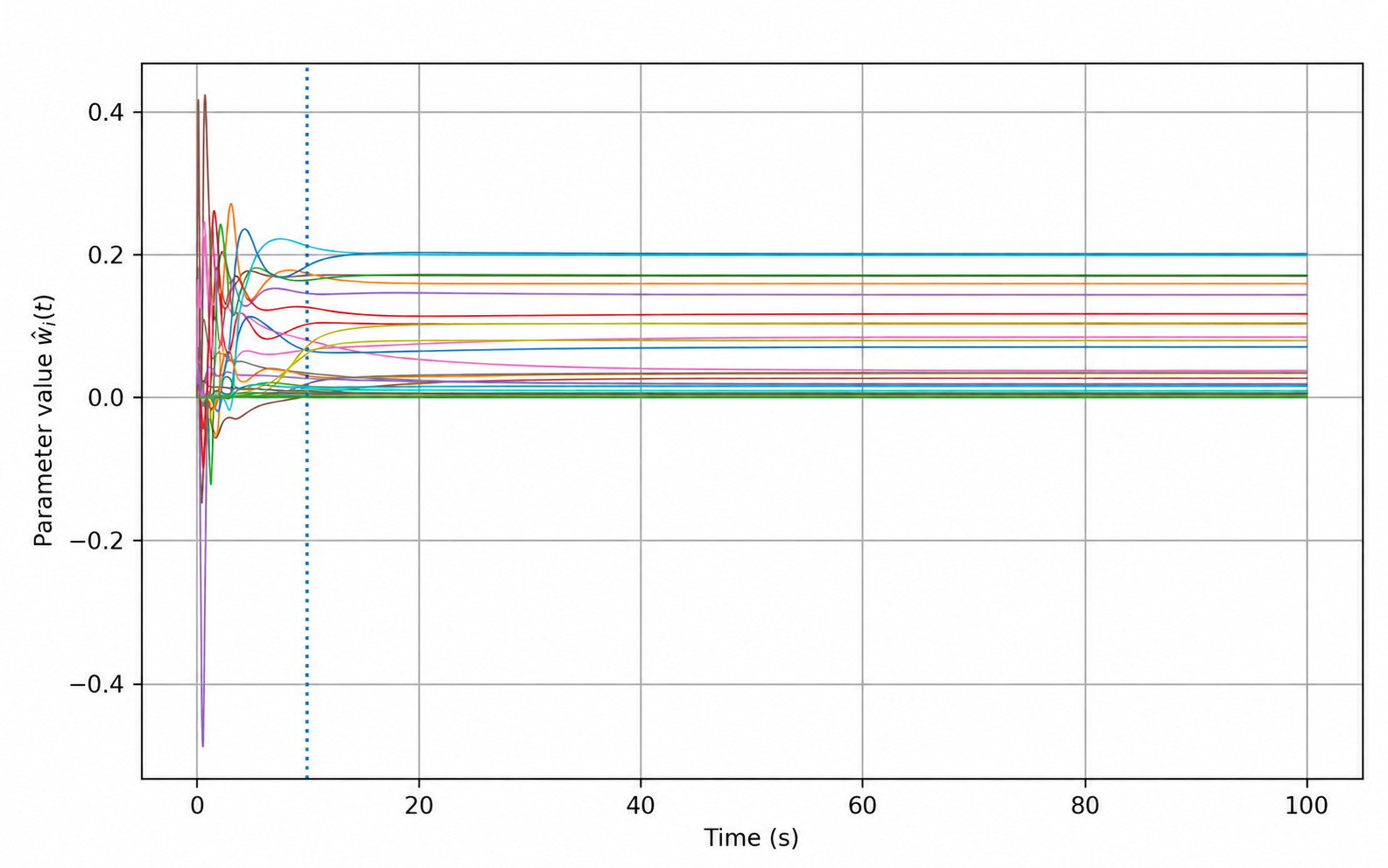}\\[-1mm]
{(a) EDL: RBF weight evolution}
\end{minipage}\hfill
\begin{minipage}{0.31\textwidth}
\centering
\includegraphics[width=\linewidth]{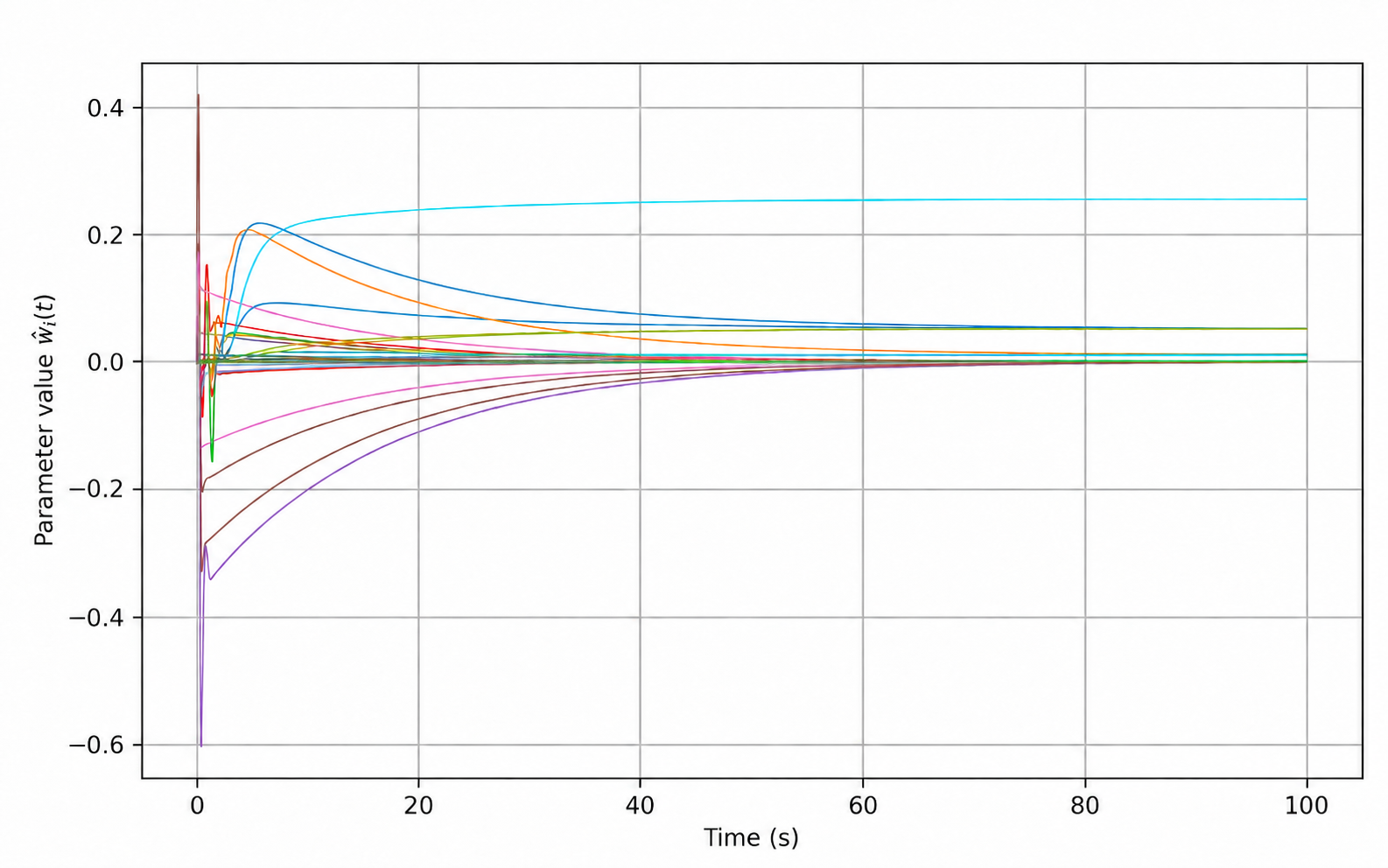}\\[-1mm]
{(b) CDL: RBF weight evolution}
\end{minipage}\hfill
\begin{minipage}{0.38\textwidth}
\centering
\includegraphics[width=\linewidth]{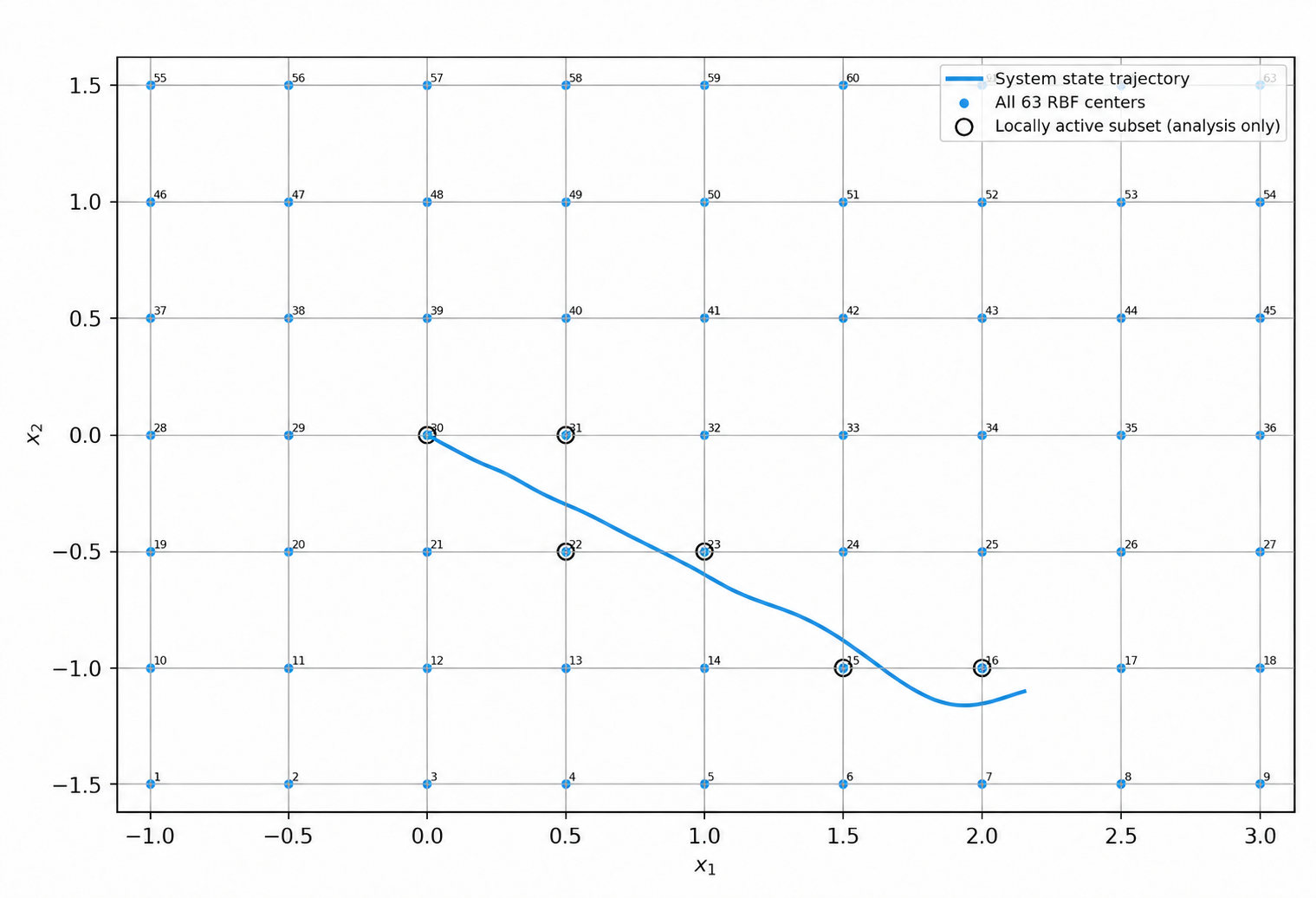}\\[-1mm]
{(c) Nonrecurrent state-space trajectory}
\end{minipage}\\[1.5mm]
\begin{minipage}{0.4\textwidth}
\centering
\includegraphics[width=\linewidth]
{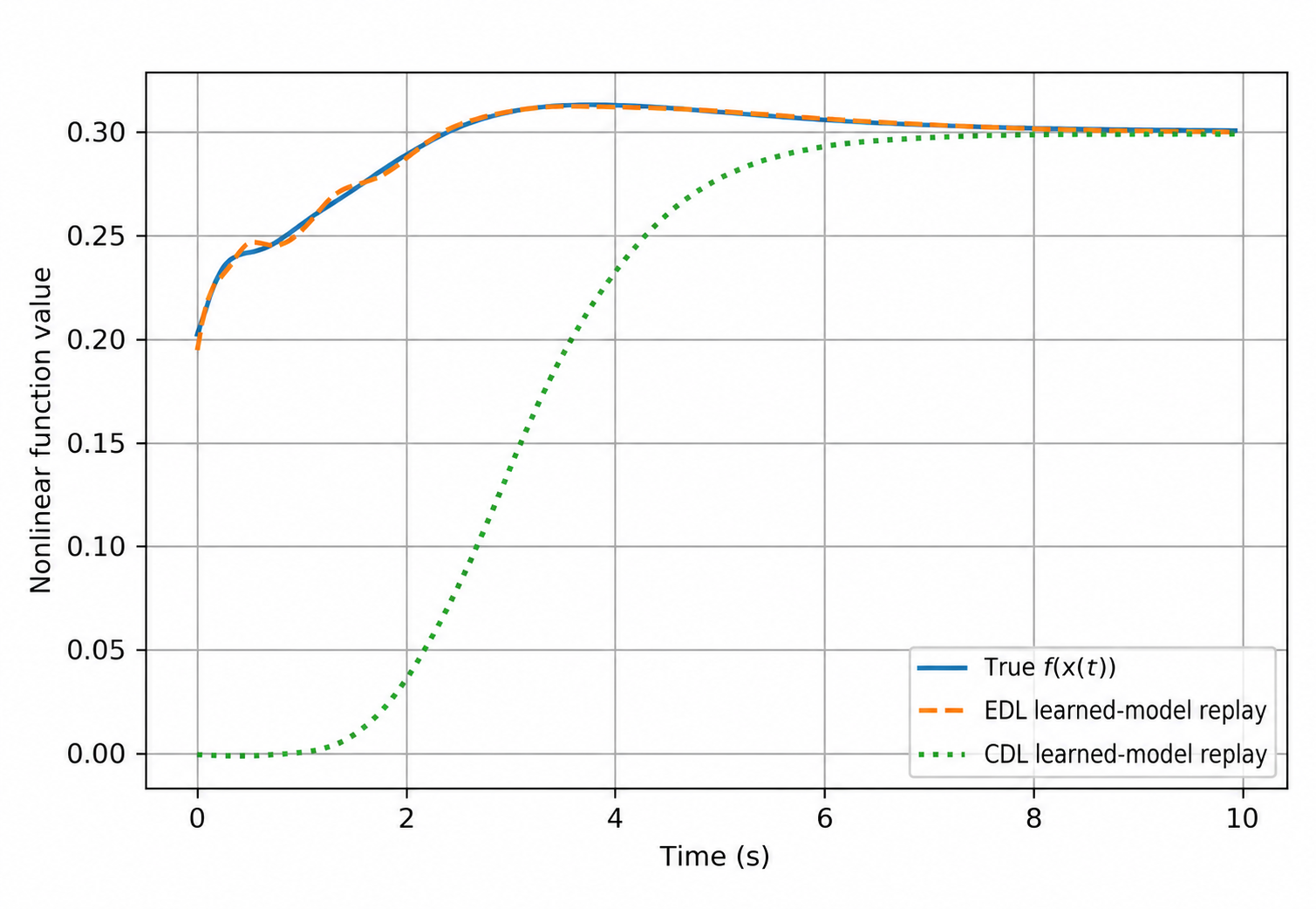}\\[-1mm]
{(d) Unknown function approximation}
\end{minipage}\hfill
\begin{minipage}{0.6\textwidth}
\centering
\includegraphics[width=\linewidth]{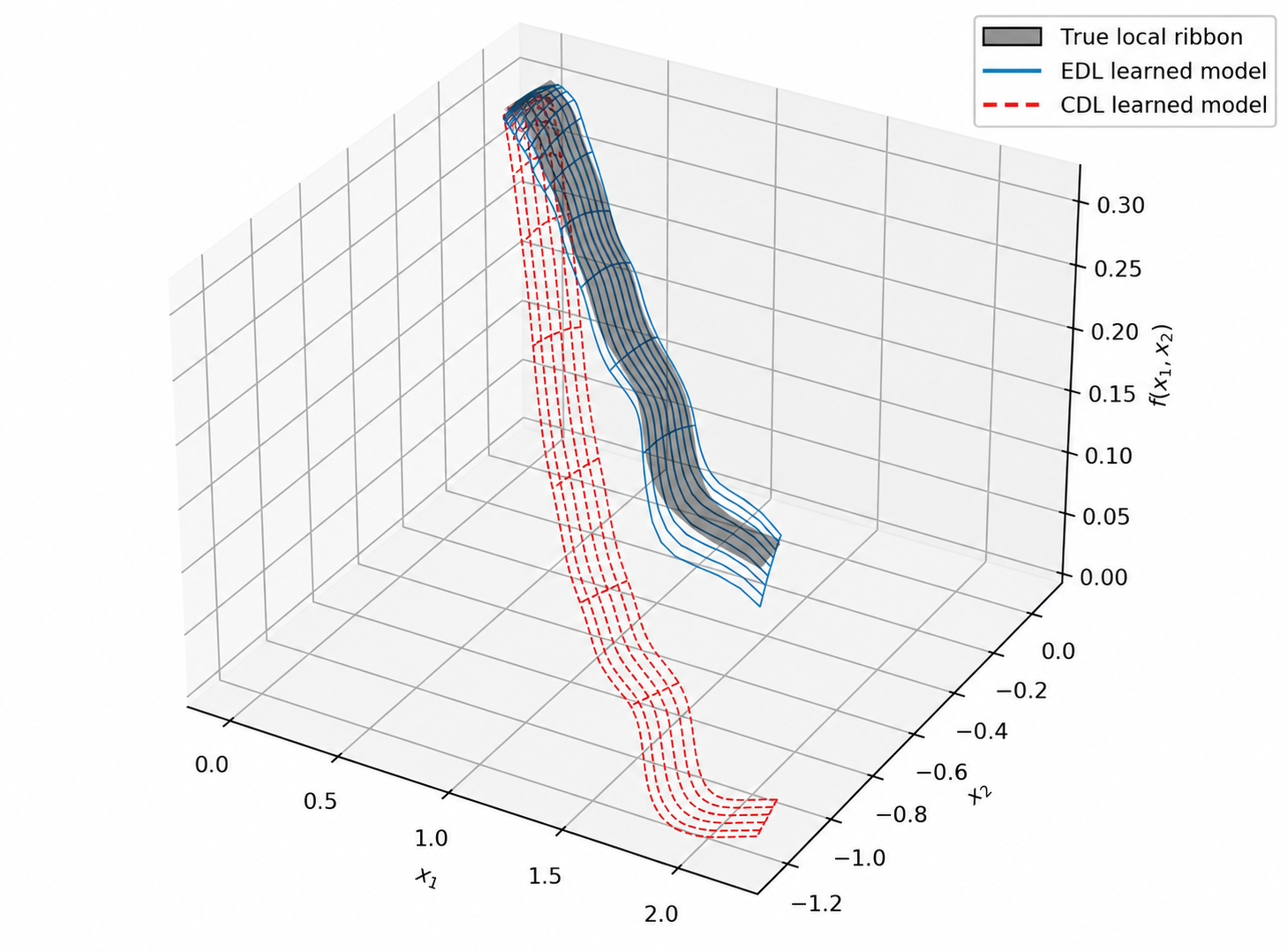}\\[-1mm]
{(e) Local approximation in state space}
\end{minipage}
\caption{Learning results. (a) and (b) show all $63$ neural-network weights for EDL and CDL, respectively. The dotted line in (a) marks the automatically selected EDL freezing time $t_b=9.927$ s; CDL in (b) runs continuously and has no freezing operation. (c) shows the finite state-space trajectory together with all uniformly distributed Gaussian RBF centers; the six local neurons used in the local-FTE check are highlighted. (d) compares the approximation quality of the resulting final EDL and CDL neural models against the true model. (e) gives a three-dimensional local-ribbon visualization around the experienced trajectory, where the gray ribbon is the true local nonlinear function and the blue and red curves show the learned EDL and CDL representations, respectively.}
\label{fig:all-weights}
\end{figure*}

The regulation trajectory passes through only a small portion of the uniformly distributed RBF network, as shown in Fig.~\ref{fig:all-weights}(c). Specifically, the local RBFs whose centers lie along the experienced transient and are appreciably activated form the subset
\begin{equation*} \zeta=\{15,16,22,23,30,31\}.
\end{equation*}
For this subset,
\begin{equation*} \lambda_{\min}(M_{\zeta\zeta})=5.19\times10^{-3}>0,
\end{equation*}
whereas the complete information matrix is numerically rank deficient,
\begin{equation*} \lambda_{\min}(M)\approx0.
\end{equation*}
These values demonstrate that local FTE holds without excitation of the complete 63-neuron network. All $63$ uniformly distributed Gaussian RBF neurons are used in the implementation; the subset $\zeta$ is identified after the simulation solely for illustration and is not used online.

For comparison, CDL uses the same system settings and the adaptive learning law 
\begin{equation*} \dot{\widehat W}=\Gamma\left[S(x)x^TPB-\sigma\widehat W\right].
\end{equation*}
Because the regulation trajectory is nonrecurrent, the local partial-PE condition used in CDL is not maintained after the transient. Fig.~\ref{fig:all-weights}(c) confirms that the state trajectory is a finite regulation transient rather than a recurrent orbit.

\begin{figure}[t]
\centering
\includegraphics[width=0.96\columnwidth]{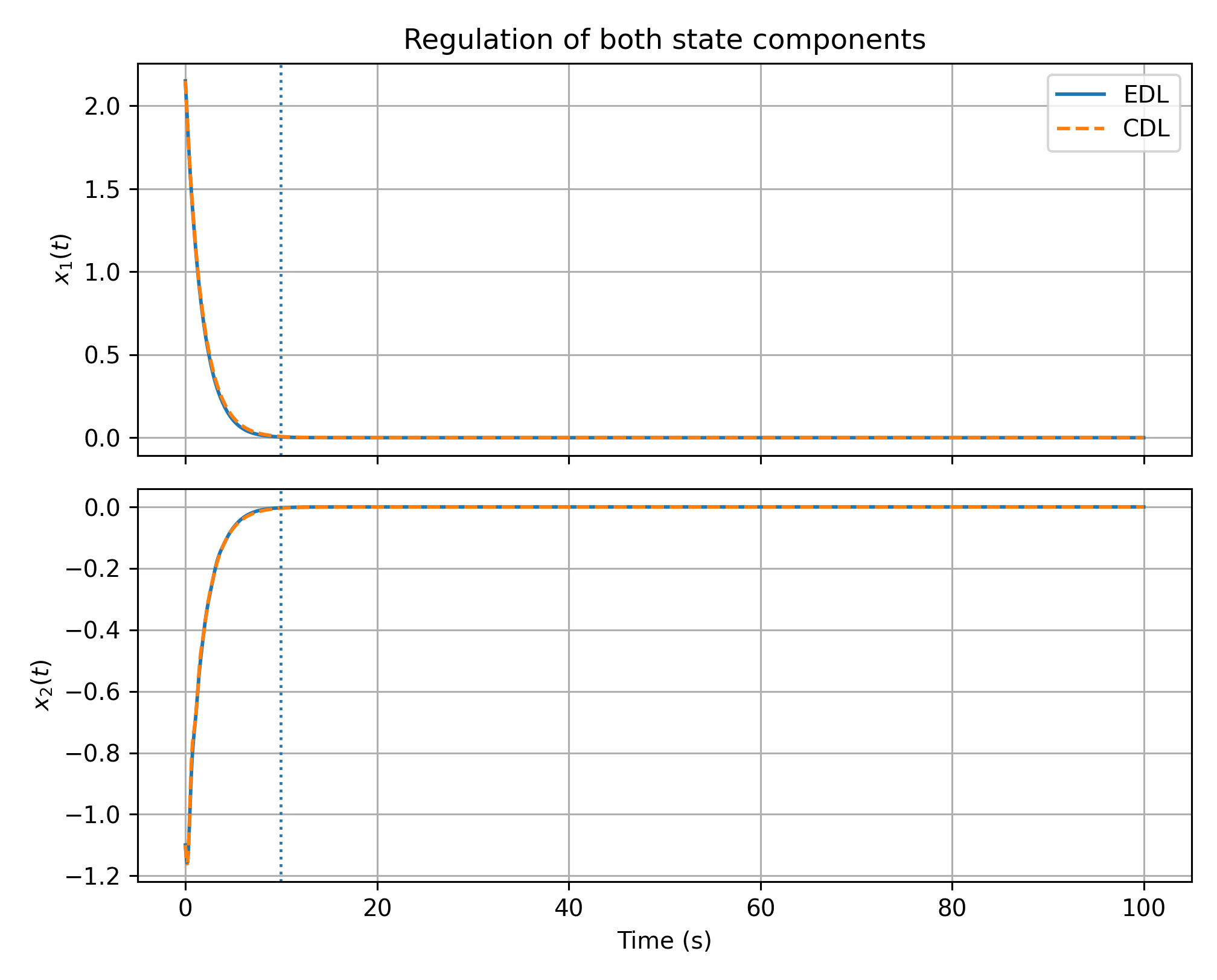}
\caption{Regulation of both state components under EDL and CDL. Both methods achieve successful regulation.}
\label{fig:new-state}
\end{figure}

Fig.~\ref{fig:new-state} compares the regulation performance of EDL and CDL. Both methods regulate the state successfully. At $t=100$ s,
\begin{equation*} \|x\|_{\rm EDL}=4.39\times10^{-4},\qquad
 \|x\|_{\rm CDL}=5.10\times10^{-4}.
\end{equation*}

The difference between the two methods is evident in the evolution of the neural-network weights. Fig.~\ref{fig:all-weights}(a) shows that, after data collection stops and the memory matrices are frozen, the EDL weights continue to evolve under the retained memory term. In contrast, CDL in Fig.~\ref{fig:all-weights}(b) has no memory and no freezing step; once the regulation transient has vanished, its parameter estimates essentially stop changing.  

Learning quality is evaluated by reconstructing the unknown nonlinear function with the final learned models, as shown in Fig.~\ref{fig:all-weights}(d)--(e). Specifically, Fig.~\ref{fig:all-weights}(d) compares the final models along the experienced trajectory. The corresponding root-mean-square (RMS) approximation errors are
\begin{equation*} e_{f,\mathrm{rms}}^{\rm EDL}=1.30\times10^{-3},\qquad
 e_{f,\mathrm{rms}}^{\rm CDL}=1.39\times10^{-1},
\end{equation*}
while the corresponding maximum errors are $7.52\times10^{-3}$ and $2.63\times10^{-1}$, respectively. This result shows that the EDL model closely reproduces the true nonlinear function along the experienced trajectory, whereas the CDL model does not. 
The learned representations are also evaluated over a narrow neighborhood of the experienced trajectory. Fig.~\ref{fig:all-weights}(e) visualizes this local ribbon in state space. On this local ribbon, the RMS surface error is $6.67\times10^{-3}$ for EDL and $1.40\times10^{-1}$ for CDL; the corresponding maximum errors are $3.38\times10^{-2}$ and $2.70\times10^{-1}$. 

This comparison illustrates the distinction between classical DL and EDL. The CDL method can regulate the plant states, but it does not preserve an accurate model of the unknown nonlinear dynamics encountered during the transient. The proposed EDL method can achieve both regulation and locally accurate learning under local FTE without requiring partial PE or persistent recurrence of the system trajectory. 

\section{Conclusion}
This paper has developed an \emph{Extended Deterministic Learning (EDL)} scheme for nonlinear regulation under local finite-time excitation (local FTE). EDL retains the central locality property of classical deterministic learning (CDL) while relaxing its excitation requirement. Whereas CDL relies on recurrent trajectories to maintain local partial PE, EDL uses memory terms in the adaptive learning law to retain sufficient local information collected over a finite interval. Continued excitation and recurrent trajectories are therefore unnecessary. The analysis establishes boundedness of all closed-loop signals, convergence of the regulation state, and exponential convergence of the locally active neural weights to small neighborhoods of their ideal values. Consequently, locally accurate identification of the nonlinear dynamics is achieved during regulation without recurrent motion. A numerical example illustrates the theoretical results and the improved local identification achieved by EDL with a nonrecurrent trajectory that does not maintain partial PE.


\begin{thebibliography}{99}
\bibitem{Narendra1990}
K. S. Narendra and K. Parthasarathy, ``Identification and control of dynamical systems using neural networks,'' \emph{IEEE Trans. Neural Netw.}, vol. 1, no. 1, pp. 4--27, Mar. 1990.

\bibitem{ParkSandberg1991}
J. Park and I. W. Sandberg, ``Universal approximation using radial-basis-function networks,'' \emph{Neural Comput.}, vol. 3, no. 2, pp. 246--257, 1991.

\bibitem{Sanner1992}
R. M. Sanner and J.-J. E. Slotine, ``Gaussian networks for direct adaptive control,'' \emph{IEEE Trans. Neural Netw.}, vol. 3, no. 6, pp. 837--863, 1992.

\bibitem{Lewis1999}
F. L. Lewis, S. Jagannathan, and A. Yesildirek, \emph{Neural Network Control of Robot Manipulators and Nonlinear Systems}. London, U.K.: Taylor \& Francis, 1999.

\bibitem{WangHill2006}
C. Wang and D. J. Hill, ``Learning from neural control,'' \emph{IEEE Trans. Neural Netw.}, vol. 17, no. 1, pp. 130--146, Jan. 2006.

\bibitem{WangHill2007}
C. Wang and D. J. Hill, ``Deterministic learning and rapid dynamical pattern recognition,'' \emph{IEEE Trans. Neural Netw.}, vol. 18, no. 3, pp. 617--630, May 2007.

\bibitem{WangChen2009}
C. Wang, T. Chen, G. Chen, and D. J. Hill, ``Deterministic learning of nonlinear dynamical systems,'' \emph{Int. J. Bifurcation Chaos}, vol. 19, no. 4, pp. 1307--1328, Apr. 2009.

\bibitem{WangHillBook}
C. Wang and D. J. Hill, \emph{Deterministic Learning Theory for Identification, Recognition, and Control}. Boca Raton, FL, USA: CRC Press, 2010.

\bibitem{ZhangWuHuWang2022}
F. Zhang, W. Wu, J. Hu, and C. Wang, ``Deterministic learning from neural control for a class of sampled-data nonlinear systems,'' \emph{Inf. Sci.}, vol. 595, pp. 159--178, May 2022.

\bibitem{ZhaoHeZhangWangHong2023}
Z. Zhao, W. He, F. Zhang, C. Wang, and K.-S. Hong, ``Deterministic learning from adaptive neural network control for a 2-DOF helicopter system with unknown backlash and model uncertainty,'' \emph{IEEE Trans. Ind. Electron.}, vol. 70, no. 9, pp. 9379--9389, Sep. 2023.

\bibitem{YangZhangWang2023}
Q. Yang, F. Zhang, and C. Wang, ``Deterministic learning-based neural control for output-constrained strict-feedback nonlinear systems,'' \emph{ISA Trans.}, vol. 138, pp. 384--396, Jul. 2023.

\bibitem{FeiLiLiLi2023}
Y. Fei, D. Li, Y. Li, and J. Li, ``Deterministic learning-based neural network control with adaptive phase compensation,'' \emph{Neural Netw.}, vol. 160, pp. 175--191, Mar. 2023.

\bibitem{WuHuZhuZhangXuWang2024}
W. Wu, J. Hu, Z. Zhu, F. Zhang, J. Xu, and C. Wang, ``Deterministic learning-based neural identification and knowledge fusion,'' \emph{Neural Netw.}, vol. 169, pp. 165--180, Jan. 2024.

\bibitem{YangZhangWang2024}
Q. Yang, F. Zhang, and C. Wang, ``Deterministic learning-based neural PID control for nonlinear robotic systems,'' \emph{IEEE/CAA J. Autom. Sinica}, vol. 11, no. 5, pp. 1227--1238, May 2024.

\bibitem{YangZhangWang2025}
Q. Yang, F. Zhang, and C. Wang, ``Knowledge fusion-based neural network control for uncertain nonlinear systems via deterministic learning,'' \emph{Int. J. Robust Nonlinear Control}, vol. 35, no. 13, pp. 5468--5487, Sep. 2025.

\bibitem{YangZhangWang2026}
Q. Yang, F. Zhang, and C. Wang, ``New results on dynamic modeling and neural control for unknown MIMO nonlinear systems via deterministic learning,'' \emph{Sci. China Inf. Sci.}, vol. 69, Art. no. 172206, Jul. 2026.

\bibitem{HuLiuWang2026}
Q. Hu, W. Liu, and C. Wang, ``Deterministic learning-based adaptive neural tracking control for nonlinear systems subject to external disturbances,'' \emph{Int. J. Robust Nonlinear Control}, pp. 1--15, 2026, doi: 10.1002/rnc.70636.

\bibitem{Chowdhary2010}
G. Chowdhary and E. N. Johnson, ``Concurrent learning for convergence in adaptive control without persistency of excitation,'' in \emph{Proc. 49th IEEE Conf. Decision Control}, 2010, pp. 3674--3679.

\bibitem{PanYu2016}
Y. Pan and H. Yu, ``Composite learning from adaptive dynamic surface control,'' \emph{IEEE Trans. Autom. Control}, vol. 61, no. 9, pp. 2603--2609, Sep. 2016.

\bibitem{PanYu2018}
Y. Pan and H. Yu, ``Composite learning robot control with guaranteed parameter convergence,'' \emph{Automatica}, vol. 89, pp. 398--406, Mar. 2018.

\bibitem{Cho2018}
N. Cho, H.-S. Shin, Y. Kim, and A. Tsourdos, ``Composite model reference adaptive control with parameter convergence under finite excitation,'' \emph{IEEE Trans. Autom. Control}, vol. 63, no. 3, pp. 811--818, Mar. 2018.

\bibitem{Parikh2019}
A. Parikh, R. Kamalapurkar, and W. E. Dixon, ``Integral concurrent learning: Adaptive control with parameter convergence using finite excitation,'' \emph{Int. J. Adapt. Control Signal Process.}, vol. 33, no. 12, pp. 1775--1787, Dec. 2019.

\bibitem{Yuan2021}
C. Yuan, P. Stegagno, H. He, and W. Ren, ``Cooperative adaptive containment control with parameter convergence via cooperative finite-time excitation,'' \emph{IEEE Trans. Autom. Control}, vol. 66, no. 11, pp. 5612--5618, Nov. 2021.

\bibitem{JiangWangMiaoNaZhaoYang2022}
Y. Jiang, Y. Wang, Z. Miao, J. Na, Z. Zhao, and C. Yang, ``Composite-learning-based adaptive neural control for dual-arm robots with relative motion,'' \emph{IEEE Trans. Neural Netw. Learn. Syst.}, vol. 33, no. 3, pp. 1010--1021, Mar. 2022.

\bibitem{XieWangDaiJiaZhang2023}
J. Xie, S. Wang, H. Dai, J. Jia, and H. Zhang, ``Adaptive neural control with fast approximation for uncertain nonlinear systems: A novel composite learning approach,'' \emph{Asian J. Control}, vol. 25, no. 6, pp. 4481--4498, Nov. 2023.

\bibitem{ShiLiYuPan2024}
T. Shi, W. Li, H. Yu, and Y. Pan, ``Efficient composite learning robot control under partial interval excitation,'' in \emph{Proc. IEEE Int. Conf. Robot. Autom.}, Yokohama, Japan, May 2024, pp. 21--26.

\bibitem{WuZouBuZhangMa2025}
Z. Wu, Z. Zou, X. Bu, J. Zhang, and K. Ma, ``Fixed-time neural network composite learning control for uncertain nonlinear systems,'' \emph{Eng. Appl. Artif. Intell.}, vol. 141, Art. no. 109722, 2025.
\end{thebibliography}
\end{document}